# Scanning cavity OF-CEAS technique for rapid collection of high-resolution spectra

Christopher A. Curwen,[1] Mathieu Fradet,[1] and Ryan M. Briggs[1]

[1]Jet Propulsion Laboratory, California Institute of Technology, Pasadena, CA 91109, USA

**Abstract:** We present a modified approach to laser optical-feedback cavity-enhanced absorption spectroscopy. The technique involves continuously scanning the length of a high-finesse cavity to periodically lock a diode laser to the cavity resonance, resulting in a discrete set of transmission measurements that are evenly spaced in frequency. For a fixed laser bias, data can be collected spanning a spectral bandwidth equivalent to the free-spectral range of the cavity, with spectral resolution inversely proportional to the distance from the laser to cavity. The center frequency of this scan can be tuned by tuning the free-running laser frequency. We demonstrate the concept using a fiber-coupled 1578-nm laser and a scanning Fabry-Perot cavity to measure a series of weak $CO_2$ absorption lines with a frequency resolution of 15.6 MHz and a noise equivalent absorption coefficient of $10^{-7}$ $cm^{-1}$, limited by the moderate finesse (~5000) and short length (~5 cm) of the cavity. Individual $CO_2$ line shapes can be measured with high resolution in a single scan that takes 67 ms. The approach has a combination of characteristics that are advantageous for *in situ* instruments, such as small size, high spectral resolution, fast data collection, and minimal components.

## 1. Introduction

Cavity Enhanced Absorption Spectroscopy (CEAS) and Cavity Ring-Down Spectroscopy (CRDS) are techniques for increasing the sensitivity of laser absorption spectrometers by exploiting the long effective pathlength of a high-finesse resonant optical cavity. A variety of approaches have been developed with varying trade-offs between sensitivity, complexity, spectral resolution, and measurement time. With a fixed cavity, data can only be collected at a discrete set of spectral points determined by the free-spectral range of the cavity (FSR = $c/2L_c$, where $L_c$ is the physical length of the cavity). Cavities on the scale of a meter are typically required to resolve molecular line shapes. Techniques such as Optical-Feedback (OF-) CEAS [1, 2], continuous-wave (cw-) CRDS [3, 4], and Frequency-Agile Rapid Scanning (FARS) spectroscopy [5] can be used to rapidly collect such data. The same techniques can be used to collect higher spectral resolution data by stepping the length of the cavity to tune the resonant frequencies, requiring some stabilization time in between [6-9]. Pound-Drever-Hall locking loops can achieve the highest sensitivities and continuous spectral data, but are arguably the most complicated techniques [10-12].

In this manuscript, we present a modified version of OF-CEAS that can rapidly measure high-resolution spectra with a short cavity and a minimal setup that does not require any locking or feedback loops. We use a piezoelectric actuator to scan the length of the high-finesse cavity, periodically locking a distributed feedback (DFB) diode laser to a set of evenly spaced discrete frequencies with spacing inversely proportional to the propagation distance from the cavity to the laser. The technique can be enabling for *in situ* instruments that require small size and minimal complexity. We present theoretical descriptions and simulations of the concept as well as an experimental demonstration measuring a series of $CO_2$ overtones around 1578 nm.

## 2. Description of technique

The experimental configuration is illustrated in Fig. 1. The output of the single-frequency diode laser (Norcada NL1578-B) is coupled to a polarization maintaining (PM) fiber and sent from port 2 to port 3 of a fiber circulator (Thorlabs CIR1550PM-APC). The signal is then coupled to free-space via a GRIN fiber collimator (Thorlabs 50-1550A-APC), and coupled to a tunable FP cavity (Thorlabs SA30-144) using a 60 mm focal length lens. The transmitted signal is then coupled back to PM fiber using a symmetric lens and coupler configuration, and passed through a variable fiber attenuator (Thorlabs VOA50PM-APC) before feeding back to the laser via port 1 of the circulator. A fiber coupler is used to split-off most of the power transmitted through the FP cavity to a photodetector, which is then amplified (Thorlabs SA201) and sent to a DAQ board (NI USB-6366) for recording the signal. The circulator prevents feedback from the initial reflection at the first mirror of the FP cavity. A V-cavity could be used instead, removing the need for a circulator, but for demonstration of the technique near 1578 nm, the high-finesse FP cavity and fiber circulator are readily available. In this section, we first provide a qualitative overview of the operating principle of the setup, and then we will provide a more detailed analysis.

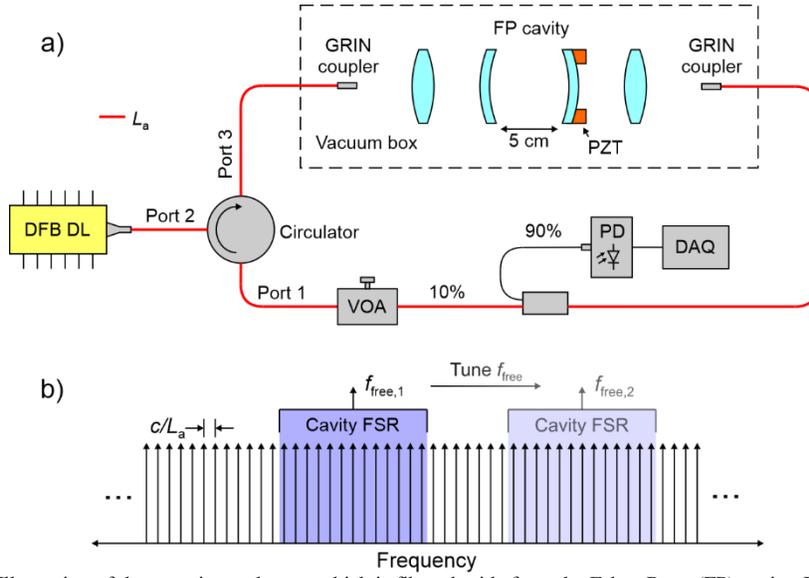

**Fig. 1.** a) Illustration of the experimental setup which is fibered aside from the Fabry-Perot (FP) cavity. DFB DL is a distributed feedback diode laser, VOA is a variable optical attenuator, and PD is the photodiode used to detect the signal. The free-space optics and fiber-to-free-space couplers are contained in a vacuum box that can be filled with the sample trace gas. b) Illustration of the operating principle. The propagation length through the fiber ($L_a$, indicated in red in (a)) establish a set of frequencies evenly spaced by $c/L_a$ [Hz] at which the laser can lock. With a fixed free-runniung laser frequency and a scanning cavity, the setup will lock to all of the modes within an FSR of that free-running frequency, as indicated by the blue shaded region. This center frequency of this scan range can be tuned by tuning the free-running frequency.

### 2.1 Qualitative description

In order for a laser to lock to optical feedback from a high-finesse cavity, a specific feedback phase is required depending on the exact laser design and bias. Typically, setups for locking a laser to a cavity include a piezoelectric mirror that can adjust the optical path length from the laser to the cavity, $L_a$, adjusting the phase accumulated in the round trip to maintain lock. Alternatively, in our setup, $L_a$ is fixed, and the feedback phase can be varied by changing the length (resonant frequency) of the FP cavity. The FP cavity phase can vary by a full $2\pi$, so scanning the length of the cavity ensures that the ideal phase for locking can be accessed. In fact, as detailed below, the proper phase for feedback occurs periodically when the cavity length changes by $\Delta L_c = \lambda_0 \times (L_c/L_a)$, and the corresponding locked laser frequency changes in steps of $\Delta f = c/L_a$. This frequency spacing corresponds to the fact that the phase accumulation over a fixed distance, $L_a$, varies by $2\pi$ with a frequency change of $c/L_a$. Therefore, by scanning the length of the FP cavity, we can collect a series of data points spaced in frequency by $c/L_a$.

Because $L_a$ is guided through fiber, it is straightforward to increase this distance and make the spectral resolution much smaller than the FSR of the cavity. One transmission measurement is collected per frequency, in series, over the course of a single cavity scan. No thermal or mechanical stabilization time is needed between data points, the free-running laser frequency is fixed while the cavity is continuously sweeping. The rate at which the scan can be performed is limited only by the laser locking time as set by the cavity lifetime and laser dynamics. A high-resolution spectrum can be collected very quickly in a single cavity scan, or multiple scans can be averaged to reduce noise. As long as the free-running laser frequency is fixed, however, data collection is limited to a bandwidth equal to the cavity FSR, centered around the free-running laser frequency. This is because the laser will always lock to the closest FP cavity mode, and there is always a mode within one FSR of the free-running laser frequency. To collect data over a different spectral range, the free-running laser frequency must be tuned. This is illustrated in Fig. 1(b). The fixed $L_a$ establishes a set of evenly spaced locked modes spanning the frequency axis, and the portion of the set that is sampled by scanning the length of the cavity is determined by the free running laser frequency and the FSR of the cavity. There is a small offset in the exact position of the set of locked modes that is dependent on the free-running frequency of the laser (more details below).

## 2.2 Theory and simulations

The behavior of the laser with feedback from the transmission of the FP cavity is given by [13]:

$$\omega_L - \omega_{\text{free}} = -\frac{\sqrt{\beta(1+\alpha_H^2)}}{2\tau_d}\left[P(\omega_L)\sin\left(\frac{\omega_L}{c}n_a L_a + \theta\right) - Q(\omega_L)\cos\left(\frac{\omega_L}{c}n_a L_a + \theta\right)\right] \quad (1)$$

Where $\omega_L = 2\pi f_L$ is the lasing frequency, $\omega_{\text{free}} = 2\pi f_{\text{free}}$ is the natural lasing frequency without external feedback, $\tau_d = \frac{n_d L_d}{c}\frac{\sqrt{R_d}}{1-R_d}$ is the photon lifetime in the semiconductor laser ($R_d$ is the reflectivity of the laser facet, and $L_d$ is the cavity length of the laser diode with refractive effective modal index $n_d$), $L_a$ is the additional propagation length before and after the high-finesse external cavity, $n_a$ is the refractive index of the fiber, $h_f(\omega) = P(\omega) + jQ(\omega)$ is the transfer function of the external cavity, $\alpha_H$ is the linewidth enhancement factor of the laser gain material, and $\beta$ is the feedback strength. In the case of an FP cavity:

$$h_f(\omega) = \frac{(1-R_c)e^{-j\omega L_c/c}}{1 - R_c e^{-j\omega 2L_c/c}} \quad (2)$$

Where $R_c$ is the reflectivity of the FP mirrors, and $L_c$ is the length of the FP cavity. Substituting Eq. (1) into Eq. (2), we get:

$$\omega_{\text{free}} = \omega_L + \frac{\sqrt{\beta(1+\alpha_H^2)}}{2}\frac{c}{2nL_d}\frac{F_c}{F_d} \times$$
$$\left[\frac{\sin\left(\frac{\omega_L}{c}(L_a + L_c) + \theta\right) - R_c \sin\left(\frac{\omega_L}{c}(L_a - L_c) + \theta\right)}{1 + \left(\frac{2F_c}{\pi}\right)^2 \sin^2\left(\frac{\omega_L}{c}L_c\right)}\right] \quad (3)$$

Where $F_c = \frac{\pi\sqrt{R_c}}{1-R_c}$ and $F_d = \frac{\pi\sqrt{R_d}}{1-R_d}$ are the finesse of the FP cavity and laser cavity respectively, and $\theta = \tan^{-1}\alpha_H$. This is a similar expression to that derived in Morville, *et al.* [2], but applies to the transmission of a FP cavity instead of the reflection from a V-cavity.

In Fig. 2, we consider the scenario in which the free-running laser frequency, $f_{\text{free}}$, is fixed and $L_c$ is scanned over a very small distance (~2 nm). The parameters used for the simulation are $L_c = 5$ cm, $L_a = 10$ m, $R_c = 0.9995$, $L_d = 0.5$ mm, $R_d = 0.3$, and $\alpha_H = 0$. We set $\beta = 0.02$, which gives a locking range comparable to the FSR of the FP cavity (3 GHz). $f_{free}$ is set to 1578 nm. The cavity is empty, with no absorption. In Fig. 2(a), a series of plots generated using Eq. (3) show the evolution of the laser frequency and transmitted power as the cavity scans over a distance of 7.85 nm. In each plot, the solid curve shows the relationship between the laser's free-running frequency, $f_{free}$, and the lasing frequency, $f_L$, when subject to feedback from the cavity. The frequency axes in Fig 2(a) are plotted relative to the center wavelength of 1578 nm. For each curve, there is a large range of $f_{free}$ over which $f_L$ is essentially constant as a result of the rapidly changing phase of the FP cavity with frequency, compensating for the changing refractive index of the laser ridge that would otherwise be tuning $f_{free}$. This is the basic principle of locking a laser to an external high-finesse cavity, and results in significant narrowing of the laser linewidth, stabilization of the frequency, and high transmitted power through the cavity. As the cavity length sweeps, there is an oscillation in the phase of the curve and a downward frequency drift in the resonance as a result of the cavity getting longer. If we consider the fixed free-running laser frequency indicated by the dashed lines in Fig. 2(a), then we see that the laser is locked to the cavity at points 1 and 6-10 (marked by red dots in the figure), and not locked to the cavity at points 2-5. The corresponding relative $f_L$ and transmitted power are plotted in Figs. 2 (b) and (c). When the laser is not locked, $f_L = f_{\text{free}}$ and no signal transmits through the cavity. When the laser is fully locked to the cavity, all of the signal is transmitted (when transmission equals 1). This behavior is periodic with a cavity length change of $\Delta L_c = \lambda_0 \times (L_c/L_a)$ and a frequency shift of $c/L_a$, as can be observed in Figs. 2(b) and (c). This is precisely the set of modes illustrated in Fig. 1(b).

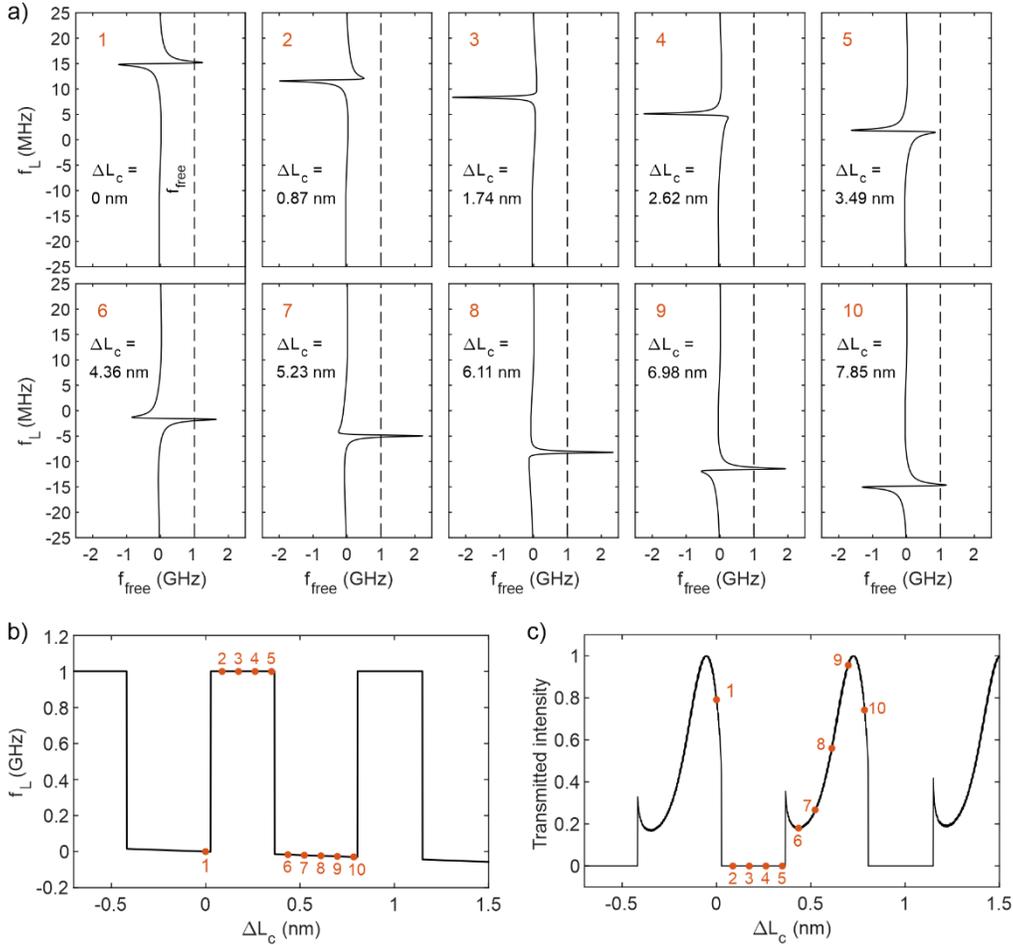

**Fig. 2**. a) Locking curves FP cavity cavity length is canned over a distance of ~ 2 nm ($L_d = 0.5$ mm, $R_d = 0.3$, $L_c = 5$ cm, $L_a = 10$ m, $R_c = 0.9995$, $\beta = 0.02$). Both the x- and y- frequency axes are relative to the center wavelength of 1578 nm. The dashed line in each figure indicates the specific free-running frequency of the laser in this test case. The laser is locked to the cavity in plots 1 and 6-10, and free-running in plots 2-5. The x-axes of the figures are on the scale of the FSR of the cavity (3 GHz), while the y-axes are on the scale of the spectral resolution of the scan (30 MHz). The locking curves have a slope of 1 outside of the perterbations caused by the FP cavity, giving $f_L = f_{free}$ for plots 2-6, though the curves appear vertical in the figures due to the scale. b) Lasing frequency and c) transmitted intensity corresponding to each of the curves in (a).

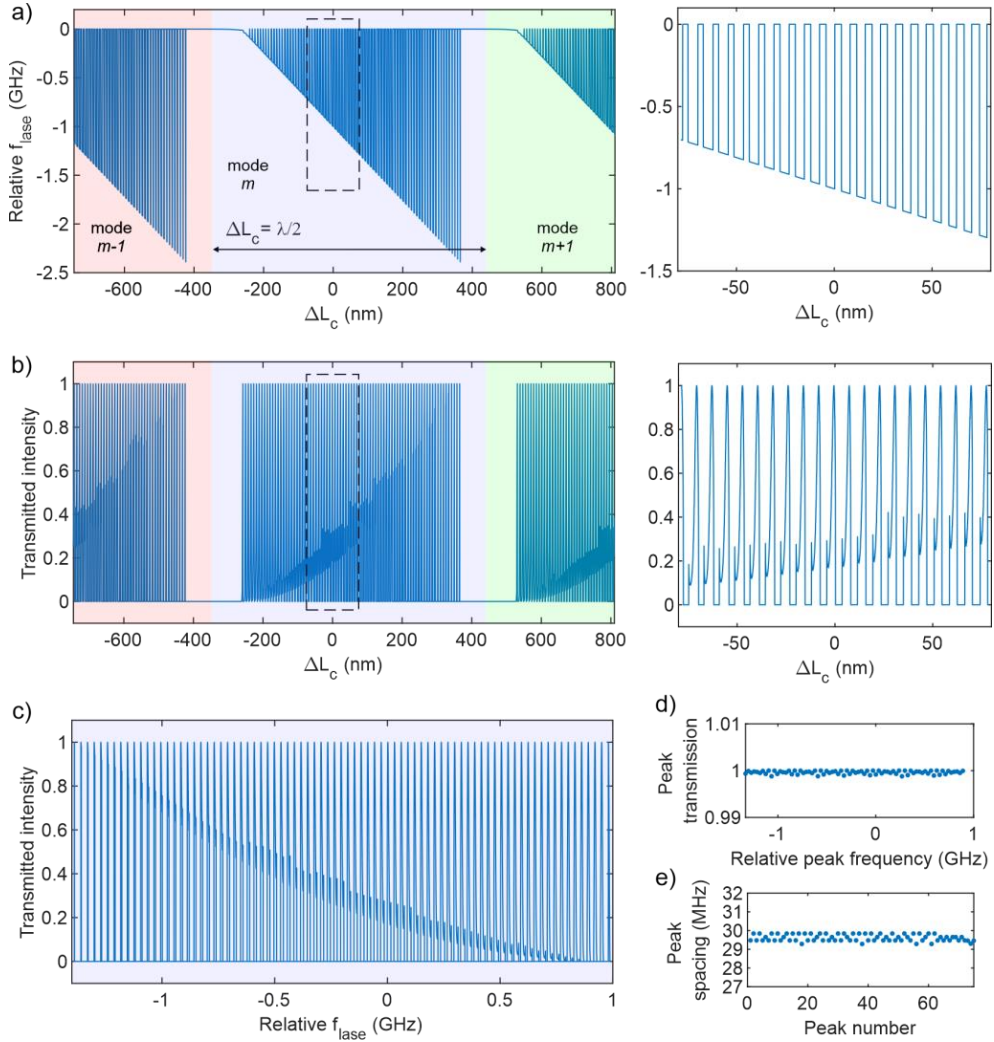

**Fig. 3**. a) Simulated relative lasing frequency and b) transmitted laser intensity for an empty FP cavity as the length (x-axis) is swept over a span of two FSRs (1578 nm). The laser locks to three neighboring modes (*m*-1, *m*, and *m*+1 where *m* is an integer) as they scan across the free-running laser frequency. The righthand plots show detailed views, as indicated by the dashed boxes. c) Simulated transmitted intensity as a function of laser frequency for the central portion of the scan. d) Peak transmission for each cycle of the curve in (c). e) Frequency spacing between peaks in (c) and (d). The cavity is empty in the simulation, so no absorption is observed and all peaks reach full transmission.

To further illustrate the behavior, in Fig. 3 we plot the simulated lasing frequency and cavity transmission over a larger scanning range of the cavity length, equal to the wavelength of the laser, 1578 nm. This corresponds to the cavity resonant frequencies scanning across a bandwidth of two FSR of the cavity. The laser locks to three neighboring modes ($m$-1, $m$, $m$+1) as they scan across the free-running laser frequency, indicated by the three shaded regions in Figs. 3(a) and (b). Because the laser frequency is fixed in this scenario while the frequencies of the cavity modes are being scanned, and the laser locks to the nearest cavity mode, these shaded regions contain the same information spanning the same frequency range. No matter how far the cavity is scanned, as long as the free-running laser frequency is fixed, one will only repeatedly collect data over the same spectral range, limited to one FSR of the cavity centered around the free-running frequency.

Focusing on center shaded portion of Fig. 3(a) (data collected as a single cavity mode scans across the free-running laser frequency), the difference between the free-running and locked frequencies is observed to become continually larger, approaching the 3 GHz FSR of the cavity, and then snaps back and the repeats as it begins locking to the neighboring cavity resonance which is then closer to the free-running frequency. We have set $\beta$ such that the locking range is slightly smaller (~2.5 GHz) than the FSR of the cavity, which results in the small portions of the scan range where no locking occurs. The laser always locks to a lower frequency due to the hysteresis of the locking process [14]. The righthand plots of Figs. 3(a) and (b) show detailed views of the center portions of the curves (indicated by the dashed boxes) so the behavior can be seen more clearly. In Fig. 3(c), the data in Figs. 3(a) and (b) are combined to plot the transmitted intensity as a function of lasing frequency, and in Fig. 3(d) and (e), the transmission and frequency spacing between peaks are plotted to demonstrate that the transmission always reaches one and the frequency spacing is constant. The frequency spacing is ~29.8 MHz, just under $c/L_a$= 30 MHz, a result of the increasing difference between the free-running and locked laser frequencies. There are minor fluctuations in the peak spacing and transmission associated with the limited temporal and spectral resolution of our simulation. Additionally, the entire set of transmission peaks can be uniformly offset in frequency by a change in the free-running frequency, but simulations indicate this offset is relatively small, limited to ~25% of the spacing between peaks (~7.5 MHz, in this case). Phase noise in the optical path of the setup will translate to noise in the frequency that a transmission peak corresponds to. The magnitude of the noise is proportional to the width of the cavity resonances, so the higher the finesse of the cavity, the lower the spectral noise.

## 3. Experimental Results

To experimentally demonstrate the technique described above, we measured a series of $CO_2$ absorption peaks from 1577 to 1580 nm, which is the full current tuning range of our DFB laser at a constant temperature of 35 °C. The FP cavity is scanned with a sawtooth pattern at a rate of 15 Hz over a distance of the wavelength, 1578 nm (two cavity FSR), giving a piezoelectric speed of 24 $\mu$m/s during data collection. With the laser bias fixed, a sample set of data recorded by during one scan of the FP cavity is plotted in Fig. 4(a). With a 30 V ramp applied to the FP cavity piezoelectric, the cavity can be scanned over ~2 FSR (6 GHz), giving a set of peaks in the middle of the scan, and the neighboring sets of peaks are visible at the edges of the scan (which contain the same information as the central peaks). This is the set of evenly spaced modes described in Figs. 1-3 (except $CO_2$ gas is present in the cavity, causing the dip in transmission over the scan). The center collection of peaks is the data collected as one cavity mode scans across the free-running laser frequency, and the data at the edges contain repeat information associated with the neighboring modes tuning across the free-running frequency. The feedback strength was selected so that the locking range is about half of the FP cavity FSR (measured $\beta = 0.0025$), though locking is observed across the full FSR with stronger feedback. The return loss of the fiber circulator is 40 dB, suppressing feedback from direct reflections far below that from the transmitted signal. The peaks between 20-30 ms in Fig. 4(a) do not achieve a strong lock as they are the most detuned from the free-running laser frequency. We discard these weakly locked peaks when processing the data. Due to the hysteresis of the locking, the cutoff on the right-hand side of the group (~45 ms) occurs when the free-running laser frequency and cavity resonance are equal, so the laser switches abruptly from a strong lock to free-running. There are a couple cycles around 45 ms that do not lock.

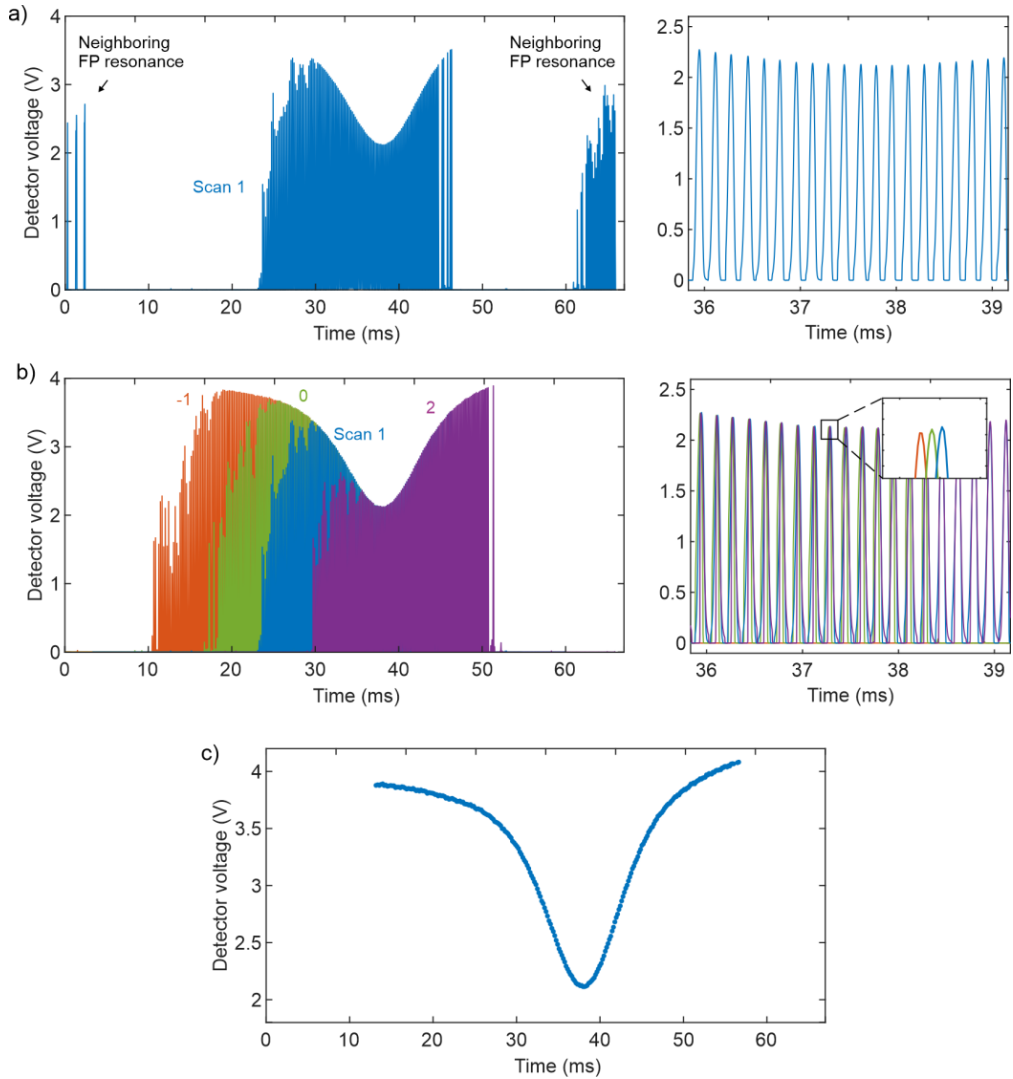

**Fig. 4.** a) Data collected during a single scan of the FP cavity with a fixed laser bias across a $CO_2$ absorption line. $CO_2$ gas was present in the cavity at a concentration of 3% by volume (other 97% $N_2$) with a total pressure of 0.13 atm. The righthand figure shows a detailed view of the center of the scan. b) Collection of four sequential piezoelectric scans, where the free-running laser frequency was tuned between scans. Inset of right-hand figure show the minor shift in the set of modes and the transmitted power as a result of tuning the free-running laser frequency and power. c) Plot of extracted transmission peaks.

The right-hand side of Fig. 4(a) shows a detailed view of the scan near the peak absorption so the behavior can be seen more clearly. There are ~100 cavity transmission peaks per scan, so ~1500 peaks are collected per second. The frequency spacing between peaks is ~15.6 MHz (calibrated from the measured lines, see below), so the 100 central peaks span ~1.5 GHz. The scan speed is limited by the time needed for the locking process to stabilize. We observe that if the scan speed is increased much beyond 15 Hz, the phase of the feedback changes too rapidly and the transmission peaks do not reach their full height. The estimated lifetime of our cavity is <1 $\mu$s (per manufacturer specification and experimental verification below), the time between peaks in a scan is ~350 $\mu$s (calculated given the FSR of the cavity, spectral resolution of the scan, and total time of the scan), and the peak of each lock spans <10 $\mu$s. The DAQ records at a rate of 600 kS/s, giving ~3 data points near each peak in cavity transmission. The bandwidth of our detector amplifier is 250 kHz. If $L_a$ were reduced, the spectral resolution would be reduced proportionally, but the scan speed could be increased because the rate of phase scanning with cavity length is proportionally slower. Additionally, we demonstrated scanning over 2 cavity FSRs in Fig. 4 for illustrative purposes, but because the two FSR contain the same information, the scan speed and data collection rate could be doubled by only scanning one FSR.

In Fig. 4(b), we plot a series of FP cavity scans where the laser bias is tuned between scans. Each subsequent scan covers a similar bandwidth, but the center frequency is slightly shifted to lower frequency corresponding to the change in the free-running frequency. The free-running frequency is changed by less than the FSR of the cavity so that there is significant overlap between scans. The peaks from the neighboring FP resonances have been omitted for clarity. Again, a detailed view of the data around the peak absorption is plotted on the righthand side of Fig. 4(b). It is observed that the peaks from all of the cavity scans are approximately aligned, though if we look even closer at the inset of the figure, each subsequent scan is actually offset very slightly from the other due to the changing free-running laser frequency, as predicted from the theory. The free-running laser frequency tunes by ~450 MHz between each cavity scan (15% of the cavity FSR), and the offset is observed to change by ~0.7 MHz (4.5% of the spacing between data points). While tuning the free-running frequency over a full FSR, the offset is observed to change by ~5 MHz (30% of the spacing between peaks), in good agreement with the simulation in Section 2. This offset dithers periodically as the free-running frequency tunes over a span of the cavity FSR, it does not accumulate as the laser continues to tune. We also note that the peaks of each subsequent scan are slightly higher in signal strength due to the increasing laser power with bias (~0.15% increase per cavity scan). To process the data into a final result, we discard the weakly locked peaks and average the overlapping peaks to obtain a single data point for each period of the locking cycle, as shown in Fig. 4(c). The offset of the modes for neighboring scans blurs the individual data points, so for maximum sensitivity, data over a single FSR should be averaged with a fixed laser bias.

In Fig. 5, we collect data over the full current tuning range of the laser. In Fig. 5(a), the blue curve shows the raw data collected under vacuum, and the red curve with $CO_2$ gas in the cavity. The ~300 GHz span of data is collected in 50 seconds, with 3 scans averaged per spectral data point. A standing wave is observed as the laser bias (frequency) is swept, which is a result of an etalon effect associated with the thickness of the FP cavity mirrors, a commonly observed phenomenon for CEAS measurements. The normalized difference between the two curves, labeled absorbance, is plotted in Fig. 5(b) and defined as $(1 - (T_{CO2}/T_{vac}))$, where $T_{vac}$ is the measured transmission through the cavity under vacuum, and $T_{CO2}$ is the transmission with $CO_2$ gas present. The measurement covers six strong $CO_2$ absorption lines, as well as several weaker lines. The approximate tuning range of the laser is known from optical spectrometer measurements, and the weak absorption lines provide confirmation of the exact lines that are measured in the (30012) ← (00001) band. There are also two resolved absorption lines from trace quantities of CO (0.1% fractional) present in the gas cylinder used for these measurements that appear near peaks 1 and 5. These CO lines appear at the expected positions relative to the $CO_2$ lines, further confirming the absolute frequency tuning range of our measurements. With these absorption lines as an absolute frequency calibration, the frequency spacing between data points is determined to be 15.6 MHz, which corresponds to a $L_a = 13.35$ m through fiber with $n_a = 1.45$ (19.35 m through free-space, approximately twice as long as the feedback length in the simulation). This is in good agreement with the known distance of fiber used.

In Figs. 5(b) and (c), the measured absorbance is plotted along with the simulated absorbance through a 5 cm FP cavity with mirror reflectance of 99.943% and $CO_2$ absorption coefficients taken from the HITRAN database [15-17]. The fitted mirror reflectance is in good agreement with the manufacturer specification of ~99.95%. While the location of the peaks match well with the simulation, the relative magnitude of the absorbance from neighboring lines does not, indicating some fluctuation in the mirror reflectance with frequency. Adjusting the mirror reflectance for each line individually, we find that the reflectivity varies from 99.940% to 99.954%, and the absorption strengths correspond to effective pathlengths between 145 and 185 m. There is a correlation between the mirror reflectivity and the alignment of the peaks relative to the ripple of the signal in Fig. 5(a) associated with the etalon effect of the mirrors (highest reflectivity when aligned with a trough in the ripple). In Fig. 6, we look more closely at one of the absorption lines (peak 2 from Fig. 5) and observe an excellent fit between the measurement and fitted simulation. The residual plot shows some rippling that may be associated with standing waves resulting from undesired reflections in the setup. The feedback path is relatively long with several fiber-to-fiber interconnects and couplers that present opportunities for stray reflections. Splicing the fiber components may help reduce these fringes.

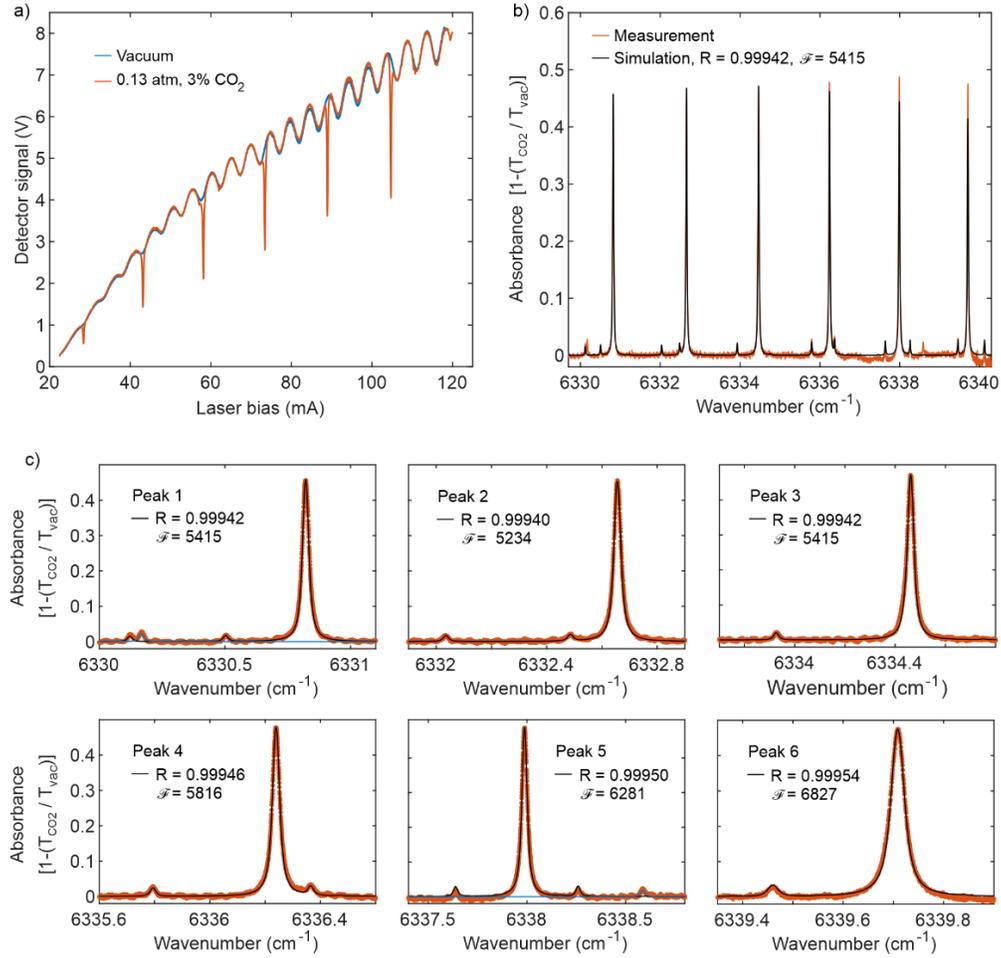

**Fig. 5.** a) Measured baseline and $CO_2$ data as a function of laser bias. b) Measured absorbance $(1 − (T_{CO2}/T_{vac}))$ compared to simulation of an FP cavity with 99.943% reflective mirrors and $CO_2$ absorption coefficients taken from HITRAN. c) Close-up on each peak with the fitted mirror reflectance and corresponding empty cavity finesse, $\mathscr{F}$, indicated. Excellent agreement in spacing between both strong and small peaks is observed. Small CO lines are also visible near peaks 1 and 5. Data is plotted in measured discrete points, but the point spacing corresponds to a spectral resolution of 0.0005 cm$^{-1}$, giving the appearance of a nearly continuous curves.

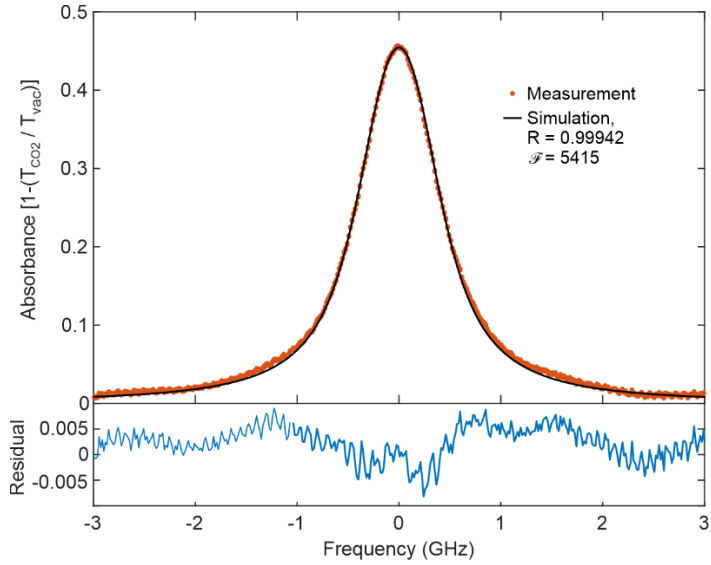

**Fig. 6.** a) Close-up on peak 2 from Fig. 5. Measurement and simulation are plotted along with the residual showing the difference between the two.

The average intensity fluctuations for each point measured for one second with an empty cavity was $\Delta I/I_{peak} \approx 10^{-3}$. Given the finesse of the cavity approximated from the fitted simulation in Figs. 5 and 6, this gives a minimum detectable absorption coefficient of ≈$10^{-7}$ cm$^{-1}$ with one second of averaging. On longer timescales, drift in the setup starts to increase noise. This is not particularly competitive with cavity enhanced systems that regularly achieve sensitivities <$10^{-10}$ cm$^{-1}$, and some cases as low as $10^{-14}$ cm$^{-1}$ [18]. We emphasize that this is meant to be a concept demonstration with available and accessible components, not an optimized system with maximum sensitivity. State-of-the-art mirrors in the near-IR and mid-IR can achieve reflectivity around two orders of magnitude higher than those used in this demonstration [19, 20], which would correspondingly improve our sensitivity by more than two orders of magnitude. Using a longer cavity, comparable to the most sensitive cavity enhanced instruments could improve our sensitivity by another order of magnitude, though enabling smaller instruments with a short cavity is part of the purpose of the presented technique, so the pros and cons must be evaluated for any given application. The ultimate sensitivity that can be achieved will depend on the stability that can be achieved to enable longer averaging times. Drift should be reduced by using more thermally stable components and through thermal stabilization of the optical fiber.

### 4. Conclusions

A modified form of OF-CEAS has been presented that involves scanning the length of a high-finesse Fabry-Perot cavity to periodically lock a semiconductor laser to the cavity, generating a set of transmission data points that are evenly spaced in frequency. The spectral spacing between data points in inversely proportional to the optical path length from the laser to the cavity, which can be made long if the setup is fibered. The technique has been described theoretically and demonstrated experimentally in a fibered 1578 nm setup with measurements of $CO_2$ absorption lines. This technique has a variety of characteristics that can make it appealing for remote *in situ* instruments, including small instrument size, high spectral resolution, a minimum number of components, no locking loops, and fast data collection rates. Thanks to the increasing availability of high reflectance mirrors and low-loss fibers throughout the mid-IR and even longwave-IR, the presented technique can likely be extended to such spectral ranges. Fiber circulators are generally not available in the mid-IR, but the circulator can be avoided by switching to a V-cavity and locking to the reflected signal. Alternatively, there are recent reports of successful OF-CEAS implementations using the reflected signal from a linear cavity [12, 21-25]. Such modified setups will be explored in subsequent studies.

**Funding.** This research was carried out, in part, at the Jet Propulsion Laboratory, California Institute of Technology, under a contract with the National Aeronautics and Space Administration.

**Disclosures.** The authors declare no conflict of interest.

**Data availability.** Data underlying the results presented in this paper are not publicly available at this time but may be obtained from the authors upon reasonable requires.

## References

1. J. Morville, D. Romanini, A. A. Kachanov, *et al.*, "Two schemes for trace detection using cavity ringdown spectroscopy," Appl. Phys. B **78** (3-4), 465-476 (2004).
2. J. Morville, S. Kassi, M. Chenevier, *et al.*, "Fast, low-noise, mode-by-mode, cavity-enhanced absorption spectroscopy by diode-laser self-locking," Appl. Phys. B **80** (8), 1027-1038 (2005).
3. D. Romanini, A. A. Kachanov, N. Sadeghi, *et al.*, "CW cavity ring down spectroscopy," Chem. Phys. Lett. **264**, 316-322 (1997).
4. G. Berden, R. Peeters, and G. Meijer, "Cavity ring-down spectroscopy: Experimental schemes and applications," Int Rev Phys Chem **19** (4), 565-607 (2000).
5. G. W. Truong, K. O. Douglass, S. E. Maxwell, *et al.*, "Frequency-agile, rapid scanning spectroscopy," Nat. Photon. **7** (7), 532-534 (2013).
6. L. Lechevallier, R. Grilli, E. Kerstel, *et al.*, "Simultaneous detection of $C_2H_6$, $CH_4$, and $\delta C^{13}$-$CH_4$ using optical feedback cavity-enhanced absorption spectroscopy in the mid-infrared region: towards application for dissolved gas measurements," Atmos. Meas. Tech. **12** (6), 3101-3109 (2019).
7. Z. F. Luo, Z. Q. Tan, and X. W. Long, "Application of Near-Infrared Optical Feedback Cavity Enhanced Absorption Spectroscopy (OF-CEAS) to the Detection of Ammonia in Exhaled Human Breath," Sensors-Basel **19** (17), (2019).
8. N. J. van Leeuwen, J. C. Diettrich, and A. C. Wilson, "Periodically locked continuous-wave cavity ringdown spectroscopy," Appl. Opt. **42** (18), 3670-3677 (2003).
9. R. Z. Martínez, M. Metsälä, O. Vaittinen, *et al.*, "Laser-locked, high-repetition-rate cavity ringdown spectrometer," J. Opt. Soc. Am. B **23** (4), 727-740 (2006).
10. R. W. P. Drever, J. L. Hall, F. V. Kowalski, *et al.*, "Laser phase and frequency stabilization using an optical-resonator," Appl. Phys. B **31** (2), 97-105 (1983).
11. J. Ye, L. S. Ma, and J. L. Hall, "Sub-doppler optical frequency reference at 1.064 μm by means of ultrasensitive cavity-enhanced frequency modulation spectroscopy of a $C_2HD$ overtone transition," Opt. Lett. **21** (13), 1000-1002 (1996).
12. G. Zhao, J. F. Tian, J. T. Hodges, *et al.*, "Frequency stabilization of a quantum cascade laser by weak resonant feedback from a Fabry-Perot cavity," Opt. Lett. **46** (13), 3057-3060 (2021).
13. J. Morville, D. Romanini, and E. Kerstel, *Cavity-Enhanced Spectroscopy and Sensing* (Springer, 2014).
14. P. Laurent, A. Clairon, and C. Breant, "Frequency noise-analysis of optically self-locked diode-lasers," IEEE J. Quantum Electron. **25** (6), 1131-1142 (1989).
15. C. S. Goldenstein, V. A. Miller, R. M. Spearrin, *et al.*, "SpectraPlot.com: Integrated spectroscopic modeling of atomic and molecular gases," J. Quant. Spectrosc. Radiat. Transfer **200**, 249-257 (2017).
16. I. E. Gordon, L. S. Rothman, R. J. Hargreaves, *et al.*, "The HITRAN2020 molecular spectroscopic database," J. Quant. Spectrosc. Radiat. Transfer **277**, (2022).
17. E. V. Karlovets, I. E. Gordon, L. S. Rothman, *et al.*, "The update of the line positions and intensities in the line list of carbon dioxide for the HITRAN2020 spectroscopic database," J. Quant. Spectrosc. Radiat. Transfer **276**, (2021).
18. J. Ye, L. S. Ma, and J. L. Hall, "Ultrastable optical frequency reference at 1.064 mu m using a C2HD molecular overtone transition," Ieee T Instrum Meas **46** (2), 178-182 (1997).
19. G. Winkler, L. W. Perner, G. W. Truong, *et al.*, "Mid-infrared interference coatings with excess optical loss below 10 ppm," Optica **8** (5), 686-696 (2021).
20. G.-W. Truong, L. W. Perner, D. M. Bailey, *et al.*, "Mid-infrared supermirrors with finesse exceeding 400,000," Nat. Commun. **14**, (2023).
21. A. G. V. Bergin, G. Hancock, G. A. D. Ritchie, *et al.*, "Linear cavity optical-feedback cavity-enhanced absorption spectroscopy with a quantum cascade laser," Opt. Lett. **38** (14), 2475-2477 (2013).
22. K. M. Manfred, L. Ciaffoni, and G. A. D. Ritchie, "Optical-feedback cavity-enhanced absorption spectroscopy in a linear cavity: model and experiments," Appl. Phys. B **120** (2), 329-339 (2015).
23. J. F. Tian, G. Zhao, A. J. Fleisher, *et al.*, "Optical feedback linear cavity enhanced absorption spectroscopy," Opt. Express **29** (17), 26831-26840 (2021).

24. F. Wan, R. Wang, H. Ge, *et al.*, "Optical feedback frequency locking: impact of directly reflected field and responding strategies," Opt. Express **32** (7), 12428-12437 (2024).
25. H. Ge, W. P. Kong, R. Wang, *et al.*, "Simple technique of coupling a diode laser into a linear power buildup cavity for Raman gas sensing," Opt. Lett. **48** (8), 2186-2189 (2023).